# Role of Dynamical Research in the Detection and Characterization of Exoplanets
A White Paper submitted to the ExoPlanets Task Force

Eric B. Ford, Fred C. Adams, Phil Armitage, B. Scott Gaudi, Mathew J. Holman,
Renu Malhotra, Geoffrey W. Marcy, Frederic A. Rasio, Steinn Sigurdsson

**1. Abstract:** Prior to discovery of extrasolar planets, the overwhelming majority of planetary research focused on explaining the properties of our own Solar System, sometimes in considerable detail. The discovery of extrasolar planetary systems revealed an unexpected diversity of planetary systems that has revolutionized planet formation theory. *A strong program of theoretical research is essential to maximize both the discovery potential and the scientific returns of future observational programs, so as to achieve a deeper understanding of the formation and evolution of planetary systems.* We outline three broad categories of theoretical research. Detailed studies of specific planetary systems can provide insights into the planet formation process (§2.1). This approach is particularly effective for multiple planet systems with strong dynamical constraints (e.g., high precision radial velocity (RV) and/or astrometric measurements). Second, theorists can test planet formation models by comparing their predictions to the observed exoplanet population (§2.2). This approach benefits greatly from wide planet surveys sensitive to a broad range of planets and stars. Finally, detailed modeling of specific physical processes is essential to understand planet formation and the origin of our solar system (§2.3). *Dynamical research plays an important role in analyzing observations for a wide range detection methods (§3) and contributes to understanding the Earth's place in the universe and the potential for Earth-like life beyond our solar system (§4).* In this white paper, we suggest how to maximize the scientific return of future exoplanet observations (§5). **Our recommendations include a strong theory program, support for multiple observational programs that will study a diverse set of planets and stars, significant observing time devoted to follow-up observations, and healthy collaboration between observers and theorists.**

**2. Relevance to Understanding the Origin, Formation & Evolution of Planetary Systems:**
A broad range of theoretical research is necessary to translate planet detections and measurement of orbits into a better understanding of the formation and evolution of planetary systems.

*2.1. Modeling specific planetary systems:* Studying the dynamical properties of specific planet systems can provide insights into planet formation processes. The detection of a single planet with unexpected properties (e.g., short orbital period of 51 Peg b, large eccentricity of 70 Vir b) can spur theoretical research to understand potential formation mechanisms (e.g., orbital migration, eccentricity excitation). Dynamical research is particularly powerful when applied to observations of multiple planet systems, since the current orbital configuration can provide clues to dynamical history of these systems. For example, detections of pairs of giant planets in mean-motion resonances provide evidence for convergent orbital migration. More detailed modeling of some of these systems has provided constraints on the eccentricity damping and/or halting of migration (Lee & Peale 2002). As another example, the secular evolution of the υ And system provides evidence for an impulsive perturbation, likely due to a previous close encounter by another planet (Malhotra 2002; Ford et al. 2005). As future planet searches uncover more multiple planet systems, we expect similar lines of research will provide insights into the dynamical history of additional planetary systems. This type of dynamical research would be most productive if combined with many high precision radial velocity and/or astrometric observations spread over a long time baseline. We encourage both RV and astrometric observatories to pursue a balanced program that includes both exoplanet searches and long-term monitoring of known exoplanet systems. Since stars hosting one giant planet are more likely to harbor additional giant planets (Wright et al. 2007), increasing the observing cadence for stars known to host at least one planet can simultaneously improve the dynamical constraints for known planets and increase the sensitivity for detecting additional planets.

*2.2. Modeling exoplanet populations:* Both the diversity of observed planetary systems and theoretical models of planet formation suggest that similar initial conditions can result in widely differing final planetary systems. While theoretical models do not predict the masses or orbits of individual systems, simplified models of planet formation can predict properties of the exoplanet population, such as the typical number of giant planets (Adams & Laughlin 2003), the inner-edge of hot-Jupiters (e.g., Ford & Rasio 2006), the eccentricity distribution (e.g., Juric & Tremaine 2007), and the correlation between stellar and planet properties (e.g., Robinson et al. 2006). Phenomenological studies can combine the observed properties of protoplanetary disks, models of planet formation and evolution, and observational selection effects to predict the properties of exoplanets discovered by various planet searches. Comparing these predictions with actual exoplanet detections (particularly wide surveys sensitive to a broad range of planets) can provide valuable constraints on models for the formation and evolution (e.g., orbital migration, eccentricity excitation) of planetary systems (Lin & Ida 2004).

*2.3. Modeling specific processes in the formation and evolution of exoplanetary systems:* Both studies of the exoplanet population and individual systems require modeling complex dynamical processes in a computationally efficient manner. A combination of analytical, semi-analytic, and numerical investigations are necessary to improve our understanding of the fundamental physical processes that are responsible for shaping planetary systems. Given the many complex, non-linear, and often chaotic processes that are essential to planet formation, it is unlikely that a first-principles simulation can include the all relevant physics. Instead, theorists will need to identify key physical processes and appropriate approximations, so that they can advance the understanding of various aspects of planetary formation and evolution. While this type of research does not depend on exoplanet observations for precise constraints, we expect that the qualitative properties of future discoveries will significantly influence the direction of such research and the phenomena that theorists seek to explain. Thus, it is important to maintain a healthy balance of observational research that helps to identify key questions and theoretical research that addresses fundamental processes in planet formation.

**3. Relevance to Detection and Characterization of Exoplanets:** Many current and future planet search techniques (e.g. RV, astrometry, pulsar timing, transit timing) are based on observing dynamical perturbations on the host stars. A planet is detected dynamically when observations are inconsistent a no-planet model, but can be explained by perturbations due to planetary companions. While the dynamical signature for a single planet is relatively simple, the dynamical signature for multiple planet systems can be much more complex and require detailed modeling. Theoretical research predicts that mature planetary systems typically contain 2 or 3 giants planets (Adams & Laughlin 2003), and RV surveys suggest that at least 30-50% of exoplanet host stars already show some evidence of additional companions (Wright et al. 2007). For systems with significant planet-planet interactions, full n-body simulations are necessary to achieve self-consistent orbital solutions (Laughlin & Chambers 2001). When early observations are consistent with multiple orbital solutions, dynamical models can identify which epochs are particularly powerful for constraining models, resulting in increased efficiency of observations (Loredo 2003). Similarly, by assuming long-term dynamical stability, theorists can reject otherwise plausible orbital solutions and constrain the masses and orbital parameters for others (e.g., Rivera & Lissauer 2000). As more multiple planet systems are discovered, dynamical research will play an increasingly important role in deciphering observations.

*3.1. Radial Velocity:* The vast majority of exoplanets discoveries are based on dynamical detections using the RV method (Butler et al. 2006). RV observations can provide sufficient constraints to enable detailed theoretical studies of multiple planet systems. As RV searches push towards higher precisions and longer survey durations, they will discover a higher fraction of multiple planet systems. *When combined with many high-precision RV observations, dynamical studies can constrain planet masses and inclinations, measure the significance of resonant and secular interactions, and provide insights into the formation and evolution of these systems.*

*3.2. Astrometry:* Astrometric planet searches could measure full 3-d orbits and planet masses without inclination degeneracies, providing strong constraints for dynamical studies. For multi-planet systems, theorists would be particularly interested in measuring relative inclinations and resonant angles. While the amplitude of RV perturbations decrease with orbital separation, the magnitude of the astrometric perturbations (or pulsar/white dwarf timing) increases with orbital separation. Therefore, astrometric planet searches could detect more distant planets, provided that astrometric observations continue for an extended time. This increased sensitivity to distant planets is also expected to increase the fraction of systems where the observed perturbations are due to multiple planets. *When combined with future astrometric observatories, dynamical studies could constrain planet masses and orbits (for planets with orbital periods exceeding the time span of observations), determine if planets are nearly coplanar, study resonant and secular interactions, and provide insights into the formation and evolution of these systems.*

*3.3. Transits:* Planets can be detected via transit photometry without appealing to dynamics. However, experience has taught that a variety of astrophysical phenomena can induce photometric variations similar to transiting planets. Therefore, it is important to confirm any putative planet detections with another method, typically by searching for the dynamical effects of a planet using RV observations (O'Donovan et al. 2006). Space-based transit searches aim to identify low-mass planets candidates and confirm them using RV follow-up observations. If these stars host additional planets (potentially more massive and not necessarily transiting), then RV follow-up would differ from that predicted based on the transit observations. In such cases, a combination of transit observations, RV observations, and dynamical modeling may enable the confirmation (or rejection) of exciting low-mass planet detections.

*3.4. Transit Timing:* If a star and one transiting planet are the only bodies in the system, then the transits will be strictly periodic. However, if an additional planet orbits the star, then the times of the transits will be affected (Miralda-Escude 2002; Holman & Murray 2005; Agol et al. 2005). By analyzing the deviations of the observed transit times from a strictly periodic model, astronomers can search for additional planets orbiting the star. The transit timing method is particularly sensitive to low-mass planets near mean motion resonances (that are difficult to detect with other methods). Since the transit timing signal will typically be dominated by the planet-planet interactions, *dynamical modeling will be needed to interpret transit timing observations in terms of planet masses and orbital parameters.* The frequency of planets in various resonances can be used to constrain theories of planet migration.

*3.5. Disk Signatures:* High resolution imaging and spectroscopy of protoplanetary and debris disks can provide constraints on the early stages of planet formation that are very complimentary to exoplanet observations. These observations can also reveal disk asymmetries most likely due to perturbations by planets. *Significant theoretical research will be necessary to interpret such observations in terms of the masses and orbits of exoplanets* (e.g., Wolf et al. 2007).

*3.6. Direct Detection:* Eventually, direct detections of extrasolar planets will enable characterization of planet atmospheres and surfaces based on spectroscopy and photometric variability. If other detection techniques have not already identified suitable planets, then direct detection campaigns would first need to devote considerable observation time to searching for such planets. Most direct detection concepts could detect a planet only when its projected separation from the host star is between an inner working angle and an outer working angle. Discovering a planet will require at least two epochs of observations to confirm that the putative planet is indeed moving with the host star, and for some designs, observations at multiple roll angles will be necessary. Therefore, either discovering a planet or obtaining a significant null result (e.g., excluding a planet larger than Earth in the habitable zone) will require several observations. *Dynamical analyses of RV, astrometric, and/or early imaging data can improve the observational efficiency by predicting times and rotation angles most favorable for detecting and excluding putative planets.* Direct detections can also provide astrometric measurements (position of the planet relative to the host star) that will enable studies of orbital properties and interactions of multiple planet systems. *Measuring a planet's orbital properties will constrain its thermal properties and aid in the interpretation of spectroscopic measurements.*

**4. Relevance to the Potential for Earth-like Life Elsewhere in the Universe:** While previous exoplanet discoveries have revealed a diverse range of planetary systems, it is not yet clear if planetary systems resembling our Solar System are common or exceedingly rare. Future observational programs will search for planets increasingly similar to the Earth, in terms of their mass, orbital separation, host star, physical size, and atmospheric/surface properties. It has been suggested that several dynamical properties of the solar system may have influenced the evolution of life on the Earth. For example, interactions between the giant planets and a massive planetesimal disk may have triggered the late-heavy bombardment of the Earth (Tsiganis et al. 2005) and contributed to the delivery of Earth's oceans (Morbidelli et al. 2000). Periodic variations in both the Earth's rotational and orbital state are believed to cause variations in Earth's climate (Hays et al. 1976). Both the small eccentricities of the solar system planets and the presence of a massive Moon that stabilizes the Earth's obliquity contribute to a stable climate that may have been significant for the evolution of life on Earth. Thus, the detection of Earth-like planets will stimulate a variety of questions about the host planetary system. "Are there signs of large scale planetary migration, such as other planets in mean-motion resonances, or giant planet at small orbital separations? Are there signs of previous violent phases of evolution, such as eccentric or highly inclined planets? Will the orbits remain nearly constant or undergo significant secular evolution? What are the implications for the planet's climate, the potential for liquid water, and the possibility of Earth-like life?" Therefore, *detections of Earth-like planets near the habitable zone should be accompanied by significant follow-up observations to identify any other planets and characterize the dynamical state and history of the entire planetary system.*

**5. Recommendations:** Despite the long history of solar system research, the study of planetary formation and evolution is still in its infancy. For example, despite the significant attention given to the "core accretion versus gravitational instability debate," neither core accretion nor gravitational instability represents a single theory that makes unique predictions. Instead, both represent broad frameworks that include significant uncertainties and adjustable parameters. While we could list specific questions that currently attract the attention of dynamicists, few are likely to withstand the test of 15 years of new observational discoveries. Historically, many exoplanet discoveries have repeatedly surprised were largely unexpected by the astronomical community (e.g., sub-Earth mass planets around the pulsar PSR B1257+12, a distant giant planet orbiting the pulsar-white dwarf binary PSR B1620-26, giant planets with orbital separations less than 1 AU, giant planets on eccentric orbits, pairs of planets participating in a mean motion resonance, planets with extremely high eccentricities and small pericenter separations, and planets with orbital periods of only 2 days). Given that *previous exoplanet discoveries have repeatedly demonstrated the limits of our understanding of planet formation theory*, our field is not yet at a stage where it is advantageous to design specific experiments to test precise predictions of one or two models. Instead, we feel that theoretical research in the formation and evolution of planetary systems will be best served by a healthy balance of new observational constraints and theoretical investigations that interpret and explain the resulting discoveries.

Between 1988 and 1992, the "Towards Other Planetary Systems" workshops and TOPS science working group produced an impressive report that recommended a three pronged approach: 1) the Keck II telescope, primarily focusing on direct detection, adaptive optics, and interferometry, 2) a space based astrometric observatory, and 3) space or lunar based large aperture interferometers to detect Earth-like planets, take their spectra, and obtain multi-pixel images of planets. As of 2007, each of these three most highly recommended approaches for detecting planets is still in it's infancy. Instead, the pulsar timing, transit photometry, and microlensing methods (only 3 of the report's 152 pages) have detected exoplanets. While the report encouraged RV searches to continue, it did not predict that this method would discover the overwhelming majority of exoplanets during the next 15 years (TOPS 1995). Fortunately, limited support of RV planet searches continued. Once this technique discovered planets using relatively modest telescopes, the general-purpose nature of the Keck observatories enabled NASA to rapidly reallocate resources to expand RV surveys. While our knowledge of giant planets has advanced considerably since 1992, we still have precious little observational constraints on Earth-like planets beyond the solar system. Thus, rather than suggest particular observational campaigns, we describe the broad conditions that are likely to maximize the scientific productivity of theoretical research in the formation and evolution of planetary systems.

## 5.1. Major Recommendations

● The 1991 and 2001 NRC decadal surveys concluded that significant funding of theoretical research is necessary to maximize the scientific return of new observatories. *Theoretical research in planetary dynamics plays an essential role in the design, analysis, and interpretation of exoplanet observations. Strong funding for theoretical research is essential to maximize the scientific return of observational programs.* The majority of dynamical and theoretical research should be funded by individual investigator grants to researchers that are not necessarily associated with any project. Individual investigator grants are critical to allow young researchers to develop innovative scientific programs. Streamlining proposal requirements and reducing the delay between submission and funding would increase the productivity of theoretical research.

● Theoretical research is most productive when guided by a steady stream of new observations. Our limited understanding of planet formation makes it difficult to choose the "best" observational programs to support for the next two decades. Ideally, theorists would like a broad portfolio of exoplanet observations, including multiple detection techniques, follow-up observations, both ground and spaced-based observatories, and a distribution of planet and host star properties. We encourage observational campaigns that have significant sensitivity to planets in previously unexplored portions of parameter space. We support the ambitious goals of detecting and characterizing Earth-like planets, yet we caution that *concentrating too great a portion of the available observational, human, and financial resources in any one observatory or method is unwise, extremely risky, and could result in retarding the advancement of the field for decades to come.* Given realistic launch dates for space missions, significant investments in ground-based facilities will be required during the coming decade to maintain the current pace of exoplanet discoveries and progress in advancing our understanding of planet formation.

● *Previous exoplanet discoveries have revealed many planets with very unexpected orbital properties.* It is worth searching for planets in any location where they could survive, regardless of the predictions of planet formation theories. Similarly, we should expect to find considerable diversity in planets' physical properties. The conventional "continuously habitable zone" (CHZ) is defined very conservatively (Kasting 1993), so planets with unexpected physical properties and/or biological processes might allow for a much broader effectively habitable zone. Searches for biosignatures should extend to planets located well outside the CHZ. Further, there is little dynamical significance to the habitable zone or one Earth-mass. Funding agencies and TACs should support theoretical research to improve our understanding of planet formation in general, even when those questions are best addressed by studying planetary systems without Earth-mass planets in the habitable zone. *It is important to search for and characterize a wide range of planets and planetary systems, including planets and host stars both similar and dissimilar to our own.* This is necessary to appreciate the significance of our Earth & Solar System.

● *Observational programs should balance the desire to study new planets with the need to obtain follow-up observations that provide precision dynamical constraints.* The power of dynamical studies increases with the number, time span, and precision of the observations for each particular planetary system. Multiple planet systems, especially those with significant planet-planet interactions, are typically much more valuable to theorists than several single planet systems for providing insights into planet formation. Once a star is determined to harbor one planet, follow-up observations should be planned to test the single-planet model and search for additional planets, whenever practical. We suggest that large observing programs making targeted observations allocate a major portion of observation time to follow-up measurements.

● *Observers should provide theorists with sufficient observational data (raw, partially and fully reduced) and other information,* so that theorists can independently determine the sensitivity of observing programs and make full use of all detections, marginal detections, and null detections.

### *5.2. Additional Recommendations*
● Given the considerable diversity of orbital properties of known exoplanet systems and physical properties of planets and moons in our solar systems, it is imperative that observational, human, and financial resources not be too focused on a small number of planets. In particular, *flagship missions to characterize Earth-like planets should be designed so that they can study at least 10 planetary systems containing an Earth-like planet.*

- If financial constraints severely limit new observational capabilities, *dynamicists should pursue exciting science, guided primarily by proven ground-based proven detection techniques.* For example, a 6m telescope dedicated to RV observations would rapidly provide a cost-effective and powerful tool for detecting low-mass exoplanets, making precision measurements of multiple planet systems, and enabling detailed dynamical studies of the evolution of planetary systems. Similarly, a geographically distributed network of many small telescopes could significantly increase the capabilities for transit, transit timing, and microlensing planet searches.

- For planetary systems containing a single detectable planet, theorists desire a qualitatively unique orbital solution, but do not require high precision measurements of planet mass and orbital parameters. Shallow and wide planet searches can provide significant statistical constraints for testing planet formation models. In the event that practical limitations prevent sufficient follow-up observations of all known exoplanets, theorists should collaborate with observers to identify the most interesting systems to be targeted for future observations.

- *Dynamical studies of multiple planet systems are much more powerful when observations provide precise measurements of planet masses and orbital parameters. Significant RV and/or astrometric follow-up observations should be planned for as many multiple planet systems as practical.* Theorists should collaborate with observers when planning and scheduling follow-up observations to maximize their utility.

- To understand the orbital evolution of a planet, *it is important to detect and characterize the orbital parameters of all major planets orbiting the host star.* For example, a detection of our solar system that only identified Earth and Jupiter would not provide enough information to understand the secular orbital evolution of the Earth. Thus, when searching for planets near the habitable zone, it is important to have significant sensitivity for detecting additional planets at distances much closer and more distant than the habitable zone. Extended time baselines and/or large outer working angle are valuable for studying long-period planets.

- *Funding agencies should consult with both observers and theorists to agree upon plans for efficiently sharing data and balanced timelines for releasing observational data.* Policies should generally promote timely public availability of observations, while providing incentives that encourage observers to invest their time in exoplanet searches. *The timely release of both ground and space-based observational data is particularly important for planning space missions with unique observational capabilities and finite lifetimes.* In some cases, the desire of observers for extended proprietary periods may need to be overridden by the greater public interest of ensuring the most efficient use of space-based observatories with finite lifetimes. Data from space missions should become publicly available soon enough that outside researchers can realistically eanalyze the data, make predictions, submit observing proposals to test those predictions, and carry out the proposed observations before the end of the mission.

- Planet searches, and particularly large observational projects, should involve at least one dynamicist to ensure that the program's design and implementation will provide valuable observational constraints. Given the long lead time of large observational projects, and especially space missions, *project teams should also provide opportunities and funding for additional theorists to join the science team shortly before the first observations/mission launch (e.g., Kepler) and to initiate theoretical research throughout the project (e.g., Hubble GO).*

- Just sustaining the current level of research will require maintaining or increasing

existing levels of funding for research in the formation and evolution of planetary systems. The rapid progress in exoplanetary science has led to a significant increase in the number of active researchers. The field includes a disproportionate number of young researchers and relatively few senior scientists near retirement. Decision makers should recognize that even inflation-adjusted budgets will result in the field losing significant talent and expertise. *As significant new observational programs begin to produce data, an increase in funding for theoretical research will be necessary to maintain a healthy balance of observational and theoretical research.*

● Dynamical investigations often require significant computational resources. Without adequate computer resources, researchers are less efficient, resulting in decreased scientific productivity and cost inefficiency. *Grant agencies should not discourage 5-10% of individual investigator grants being allocated to acquiring and maintaining significant computational resources.*

● Both observational and theoretical investigations can benefit from frequent interactions between theorists and observers. *Funding agencies should support meetings to promote discussion of recent observational and theoretical results and interactions between theorists and observers.* Testing theoretical models with observations often requires a detailed understanding of observational methods and uncertainties. Thus, meetings should provide opportunities for theorists to learn about the observational sensitives, precisions, uncertainties, systematic effects, biases, etc. that should be considered when interpreting observations.

## 6. References

Adams, F. C. and G. Laughlin (2003) "Migration and dynamical relaxation in crowded systems of giant planets" *Icarus* 163, 290-306.

Agol, E., J. Steffen, R. Sari, and W. Clarkson (2005) "On detecting terrestrial planets with timing of giant planet transits" *Monthly Notices of the Royal Astronomical Society* 359, 567-579.

Butler, R. P., and 10 colleagues (2006) "Catalog of Nearby Exoplanets" *Astrophysical Journal* 646, 505-522.

Ford, E. B., V. Lystad, and F. A. Rasio (2005) "Planet-planet scattering in the upsilon Andromedae system" *Nature* 434, 873-876.

Ford, E. B. and F. A. Rasio (2006) "On the Relation between Hot Jupiters and the Roche Limit" *Astrophysical Journal* 638, L45-L48.

Hays, J. D., J. Imbrie, and N. J. Shackleton (1976) "Variations in the Earth's Orbit: Pacemaker of the Ice Ages" *Science* 194, 1121-1132.

Holman, M. J. and N. W. Murray (2005) "The Use of Transit Timing to Detect Terrestrial-Mass Extrasolar Planets" *Science* 307, 1288-1291.

Ida, S. and D. N. C. Lin (2004) "Toward a Deterministic Model of Planetary Formation. II. The Formation and Retention of Gas Giant Planets around Stars with a Range of Metallicities" *Astrophysical Journal* 616, 567-572.

Juric, M. and S. Tremaine (2007) "The Eccentricity Distribution of Extrasolar Planets" ArXiv Astrophysics e-prints arXiv:astro-ph/0703160.

Kasting, J. F., D. P. Whitmire, and R. T. Reynolds (1993) "Habitable Zones around Main Sequence Stars" *Icarus* 101, 108-128.

Laughlin, G. and J. E. Chambers (2001) "Short-Term Dynamical Interactions among Extrasolar

Planets" *Astrophysical Journal* 551, L109-L113.

Lee, M. H. and S. J. Peale (2002) "Dynamics and Origin of the 2:1 Orbital Resonances of the GJ 876 Planets" *Astrophysical Journal* 567, 596-609.

Loredo, T. J. and D. F. Chernoff (2003) "Bayesian adaptive exploration" Statistical Challenges in Astronomy 57-70.

Miralda-Escudė, J. (2002) "Orbital Perturbations of Transiting Planets: A Possible Method to Measure Stellar Quadrupoles and to Detect Earth-Mass Planets" *Astrophysical Journal* 564, 1019-1023.

Morbidelli, A., J. Chambers, J. I. Lunine, J. M. Petit, F. Robert, G. B. Valsecchi, and K. E. Cyr (2000) "Source regions and time scales for the delivery of water to Earth" *Meteoritics and Planetary Science* 35, 1309-1320.

O'Donovan, F. T., and 10 colleagues (2006) "Rejecting Astrophysical False Positives from the TrES Transiting Planet Survey: The Example of GSC 03885-00829" *Astrophysical Journal* 644, 1237-1245.

Rivera, E. J. and J. J. Lissauer (2000) "Stability Analysis of the Planetary System Orbiting υ Andromedae" *Astrophysical Journal* 530, 454-463.

Robinson, S. E., G. Laughlin, P. Bodenheimer, and D. Fischer (2006) "Silicon and Nickel Enrichment in Planet Host Stars: Observations and Implications for the Core Accretion Theory of Planet Formation" *Astrophysical Journal* 643, 484-500.

TOPS (1995) "TOPS: Toward Other Planetary Systems. A report by the solar system exploration division" NASA STI/Recon Technical Report N 95, 28009.

Tsiganis, K., R. Gomes, A. Morbidelli, and H. F. Levison (2005) "Origin of the orbital architecture of the giant planets of the Solar System" *Nature* 435, 459-461.

Wolf, S., A. Moro-Martĭn, and G. D'Angelo (2007) "Signatures of planets in protoplanetary and debris disks" *Planetary and Space Science* 55, 569-581.

Wright, J. T., and 10 colleagues (2007) "Four New Exoplanets and Hints of Additional Substellar Companions to Exoplanet Host Stars" *Astrophysical Journal* 657, 533-545.